\documentstyle[12pt]{report}

\def\beq{\begin{equation}}
\def\eeq{\end{equation}}
\def\bea{\begin{eqnarray}}
\def\eea{\end{eqnarray}}
\def\bq{\begin{quote}}
\def\eq{\end{quote}}

\def\gappeq{\mathrel{\rlap {\raise.5ex\hbox{$>$}}
{\lower.5ex\hbox{$\sim$}}}}
\def\lappeq{\mathrel{\rlap{\raise.5ex\hbox{$<$}}
{\lower.5ex\hbox{$\sim$}}}}

\def\slash#1{\setbox0=\hbox{$#1$}#1\hskip-\wd0\hbox to\wd0{\hss\sl/\/\hss}}

\begin{document}

\begin{flushright}
MZ-TH/92-56 \\
DO-TH/92-27 \\
December 1992
\end{flushright}
\bigskip

\begin{center}
{\bf{\LARGE  {\em CP}-odd Neutral Higgs Effects in}} \\[0.45cm]
{\bf{\LARGE  {\boldmath $t\bar{t}$\unboldmath} Production}}\\[1.5cm]
\bigskip\bigskip\bigskip
{\large A.~Pilaftsis}$^{\displaystyle \, \, (a) ,}$
\footnote[1]{E-mail address: pilaftsis@vipmza.physik.uni-mainz.de}
  {\large{\em and}} \hspace{0.04cm}
{\large M.~Nowakowski}$^{\displaystyle \, \, (b) ,}$
\footnote[2]{Address after 31, Dec.~1992; Physical Research Laboratory,
Navrangpura, Ahmedabad 380 009, {\em INDIA}.}\\[0.5cm]
$^{(a)}$ Inst.~f\"ur Physik, Johannes-Gutenberg Universit\"at,
W-6500 Mainz, {\em FRG}\\[0.5cm]
$^{(b)}$ Inst. f\"ur Physik, Universit\"at Dortmund,
W-4600 Dortmund,
{\em FRG}
\end{center}

\bigskip
\bigskip\bigskip
\centerline {\bf ABSTRACT}

We study $CP$ violation in the process $e^+e^- \to t\bar{t} \nu\bar{\nu}$
at an $e^+e^-$-TeV collider. As the source of $CP$ violation we assume a
two-Higgs doublet model with an explicitly $CP$-noninvariant Higgs potential.
Sizeable $CP$-odd observables originating from the subprocess reaction,
$W^+W^- \to t\bar{t}$, may arise as a result of finite width
effects of the neutral Higgs particles. $CPT$ constraints due to
final (initial) state interactions are
also taken into account. Numerical estimates of the $CP$ asymmetry are
given.\\

\newpage

\section*{1.~Introduction}

\stepcounter{chapter}

\indent

$CP$ violation in the neutral Higgs sector with two or more Higgs doublets
has recently  received much attention due to some attractive features
of these models~[1-3]. For
instance, these models can give rise to
a large electric dipole moment ($EDM$) for the
neutron~[4] as well as for the top quark~[5,6]. Non-negligible
contributions to $EDM$s for leptons~[4,7] and gauge bosons $W^\pm$, $Z^0$~[8]
have also been found.
Last but not least, the  $CP$-violating Higgs sector has entered the domain of
cosmology, namely as a natural explanation of baryogenesis~[9]. Note here that
the strength of $CP$ violation from the Cabbibo-Kobayashi-Maskawa
matrix is not sufficient to explain the baryon asymmetry in the universe~[10].
Therefore, models manifesting $CP$ violation in the scalar sector offer an
interesting field of investigation for $CP$-odd effects outside the well
known $K^0-\bar{K}^0$ system.\\

Beside the possibility of searching for sizeable signals of $EDM$s one
can also probe these models in top production and decay~[7,11].
Here, however, we suggest a different way of testing $CP$ violation
which can be mediated by neutral Higgs bosons. By considering
the production of left- and right-handed $t\bar{t}$ pairs in a high energy
$e^+e^-$-collider (i.e.~$e^+e^-\to \nu_e \bar{\nu}_e t_\lambda
\bar{t}_\lambda,\  \lambda\lambda=LL,\ RR$), one can show that the
simultaneous presence of $CP$-even ($\bar{\psi}\psi$) and $CP$-odd
($i\bar{\psi}\gamma_5\psi$) fermionic operators in the Yukawa sector
induces a $CP$ asymmetry in the event-rate difference of $N(t_L\bar{t}_L)-
N(t_R\bar{t}_R)$~[5]. To obtain a non-zero asymmetry, one should have two
interfering amplitudes with different weak phases as well. This requirement
is effectively satisfied here by using complex Breit--Wigner $(BW)$ propagators
for resonant particles~[11], i.e.~Higgs scalars~[12].
Nevertheless, it is known that
the $CP$-even phases of final and (in our case) also initial state interactions
cannot contribute to a $CP$ asymmetry~[13,14]. This is sometimes called
the $CP-CPT$ connection~[14]. We take such $CPT$ constraints into account
in order to establish a genuine signal of $CP$ violation.\\

Our paper is organized as follows. In Section 2 we outline the basic
features of a $CP$-violating two-Higgs doublet model. In Section 3 we
calculate the subprocess $W^+W^- \to t\bar{t}$ with intermediate
Higgs bosons in the $s$-channel.
Section 4 deals with $CPT$ constraints and the gauge symmetry imposed on
our process. Finally, in Section~5 we present numerical estimates and
discussion of our results.\\

\section*{2.~The {\em CP}-violating two-Higgs doublet model}
\stepcounter{chapter}
\indent

The simplest realization of $CP$ violation in the neutral Higgs sector is
a two-Higgs doublet model where the usually imposed discrete symmetry
$D$: $\Phi_1 \to -\Phi_1$, $\Phi_2 \to \Phi_2$ is softly broken~[15].
As usual the Higgs part of the Lagrangian reads
\beq
{\cal L}_H = \sum\limits_{i=1}^{2} (D_\mu\Phi_i)^+(D^\mu\Phi_i)\ -\
V_H(\Phi_1,\Phi_2),
\eeq 
where $D_\mu$ is the covariant derivative. The potential $V_H$ can be split
into two terms according to the sum,
\beq
V_H(\Phi_1,\Phi_2) \ = \ V_D(\Phi_1, \Phi_2)\ +\ \Delta V(\Phi_1, \Phi_2).
\eeq 
$V_D$ respects the discrete symmetry $D$ and can be conventionally written in
the form~[15]
\bea
V_D(\Phi_1,\Phi_2)\  & = &\ \mu_1^2(\Phi_1^\dagger\Phi_1) \ +\
\mu_2^2(\Phi_2^\dagger\Phi_2)\ +\ \frac{1}{2}\lambda_1(\Phi_1^\dagger\Phi_1)^2
\nonumber\\
&&+\ \frac{1}{2}\lambda_2(\Phi_2^\dagger\Phi_2)^2
+\  \lambda_3(\Phi_1^\dagger\Phi_1)(\Phi_2^\dagger\Phi_2)\nonumber\\
&&+\  \lambda_4(\Phi_1^\dagger\Phi_2)(\Phi_2^\dagger\Phi_1)\ +\
\frac{1}{2}\lambda_5(\Phi_1^\dagger\Phi_2)^2\nonumber\\
&&+\  \frac{1}{2}\lambda_5^\ast(\Phi_2^\dagger\Phi_1)^2 \, ,
\eea 
whereas $\Delta V$ softly violates the $D$ symmetry by introducing linear
$\Phi_1^\dagger\Phi_2$ terms
\beq
\Delta V\ =\ \lambda_6 (\Phi_1^\dagger\Phi_2)\ + \
\lambda_6^\ast (\Phi_2^\dagger\Phi_1).
\eeq 
One can check explicitly by performing $CP$ transformation on the
$\Phi_i$'s that $V_H$ is $CP$ invariant only if
\beq
\mbox{Im}\lambda_5\lambda_6^{\ast 2}\ =\ 0
\eeq 
holds. After spontaneous symmetry breaking with two complex vacuum
expectation values ($VEV$s) $v_1$ and
$v_2$ of the fields $\Phi_1$ and $\Phi_2$, respectively,
eq.~(2.5) becomes
\beq
\mbox{Im}\lambda_5(v_1^\ast v_2)^2 \ = \ 0,
\eeq 
which in turn is equivalent to
\beq
\mbox{Im}\lambda_6(v_1^\ast v_2) \ = \ 0.
\eeq 
Note that without loss of generality we can choose $\lambda_5$ real and,
say, $v_1$ real. This follows from the fact that we have the freedom
of redefining the fields by an arbitrary phase transformation
\beq
\Phi_i \ \longrightarrow e^{i\alpha_i}\ \Phi_i.
\eeq 
It is worth mentioning that due to this phase symmetry there is no
$CP$ violation in $V_H$ if either $\lambda_5$ and $\lambda_6$ is identical
to zero. This is also evident from~(2.6) and~(2.7). Writing
the fields $\Phi_1$ and $\Phi_2$ in the form
\beq
\Phi_1\ =\ \left( \begin{array}{c}
\phi^+_1 \\ \frac{1}{\sqrt{2}} (v_1+\phi_1^0+i\chi_1^0)
\end{array} \right) ,\quad \qquad
\Phi_2\ =\ \left( \begin{array}{c}
\phi^+_2 \\ \frac{1}{\sqrt{2}} (|v_2|e^{i\xi}+\phi_2^0+i\chi_2^0)
\end{array} \right) ,
\eeq 
and collecting quadratic terms in the components of $\Phi_i$ we get, after
diagonalizing the mass matrices, the following mass eigenstates
\beq
\left( \begin{array}{c} H^+ \\ G^+ \end{array} \right)\ =\
U_H\ \left( \begin{array}{c} \phi^+_1 \\ \phi^+_2 \end{array} \right), \qquad
\quad \left( \begin{array}{c}
H^0_1 \\ H^0_2 \\ H^0_3 \\ G^0 \end{array} \right) \ =\
D^TO_H \left( \begin{array}{c}
\phi^0_1 \\ \phi^0_2 \\ \chi^0_1 \\ \chi^0_2 \end{array} \right),
\eeq 
where $H^0_i$ ($i=1,2,3$) and $H^\pm$ are the physical Higgs particles and,
$G^\pm$ and $G^0$ are charged and neutral Goldstone bosons, respectively.
Defining
\beq
\tan \beta\ =\ \frac{|v_2|}{v_1} \, ,
\eeq 
the unitary transformation matrix $U_H$ in the charged Higgs sector reads
\beq
U_H \ =\ \left( \begin{array}{cc}
-s_\beta e^{i\xi} & c_\beta e^{i\xi} \\
c_\beta & s_\beta \end{array} \right) ,
\eeq 
where $\sin x\equiv s_x$ and $\cos x\equiv c_x$.
In eq.~(2.12) the phase $\xi$ is trivial and can be removed away by
appropriately rephasing the charged mass eigenstates $H^+$, $G^+$.
The orthogonal transformation matrix in the neutral Higgs sector has been
expressed for convenience as a product of two orthogonal matrices
$D$ and $O_H$. The latter has the form
\beq
O_H \ = \ \left( \begin{array}{cccc}
1  &  0  &  0  &  0 \\
0  & c_\xi & 0 & s_\xi \\
0  & -c_\beta s_\xi &-s_\beta & c_\beta c_\xi \\
0  & -s_\beta s_\xi & c_\beta & s_\beta c_\xi
\end{array} \right).
\eeq 
The matrix $D$ acts only on the physical Higgses $H^0_i$ and can therefore
be written as
\beq
D^T \ =\ \left( \begin{array}{cc}
d & 0 \\
0 & 1 \end{array} \right),
\eeq 
where $d$ is a $3\times 3$ orthogonal matrix. This matrix can be
parametrized by three Euler angles which are determined by the Higgs
couplings $\lambda_i$. At this stage it is not necessary to do the
diagonalization in full detail and to spell out $d$ in terms of
$\lambda_i$. Instead, we mention that if condition~(2.6) or alternatively~(2.7)
holds, then we get the matrix form
\beq
d\ =\ \left( \begin{array}{cc}
d' & 0 \\
0  & 1 \end{array} \right),
\eeq 
where $d'$ is a $2\times 2$ dimensional matrix. In this case, $H^0_3$
decouples from the other Higgses and is of purely pseudoscalar nature in
a $CP$-conserving two-Higgs doublet model. In general, as a consequence
of $CP$-violating terms in $V_H$, the physical neutral Higgses are mixtures
of $CP$-even and $CP$-odd components. It is then obvious that we have
\beq
d_{13}, \quad d_{23}, \quad d_{31},\quad d_{32} \quad \propto
\quad \mbox{Im}\lambda_5 (v_1^\ast v_2)^2 .
\eeq 
Among the various vertices that
are present in the full Lagrangian, we will only need
the gauge-boson interactions with Higgses, Goldstone bosons $G^\pm$ and
ghosts $c^\pm$, $\bar{c}^\pm$,
as well as the Yukawa interactions for the top quark.
It is straightforward to obtain the vertices from the
Lagrangian~(2.1) (considering also gauge-fixing terms)
\bea
V^{\mu \nu}_{WWH_i} \ & = &\ ig_WM_W\ g^{\mu \nu} (d_{1i} c_\beta
\ +\ d_{2i}s_\beta ), \\
V^{\mu \nu}_{ZZH_i}\ &=& \ ig_W\frac{M^2_Z}{M_W}\ g^{\mu \nu}  (d_{1i} c_\beta
\ +\ d_{2i}s_\beta ), \\
V^\mu_{ZH_iH_j}\ &=&\ g_W\frac{M_Z}{2M_W}\ (k_{H_j}-k_{H_i})^\mu\
d_{3i}(-s_\beta d_{1j}\ +\ c_\beta d_{2j} ) , \\
V^{\mu}_{W^\pm G^\mp H_i} \ & = &\ \pm i\frac{1}{2}\ g_W\
(k_{G}-k_{H_i})^\mu\
(d_{1i} c_\beta\ +\ d_{2i}s_\beta ), \\
V_{c^+\bar{c}^-H_i} \ & = &\ -i\frac{1}{2}\ \xi_Wg_WM_W\ (d_{1i} c_\beta
\ +\ d_{2i}s_\beta ) ,
\eea 
with the convention that all momenta are incoming and the definition
\beq
v^2 \ \ = \ \ |v_1|^2 \ +\ |v_2|^2.
\eeq 
In eq.~(2.21) $\xi_W$ denotes the gauge parameter.
In order to avoid flavor neutral currents at tree level, one has to
require the Yukawa Lagrangian $L_Y$ to be invariant under the discrete
symmetry~$D$. We choose the coupling scheme where $u$-type quarks couple to
$\Phi_2^0$ field via
\beq
L_Y\ \ =\ \  -\ \frac{1}{v_2}(\bar{U}_R M_U U_L) \Phi_2^0 \ \ + \ \ h.c.\ \ +\
\   \dots ,
\eeq 
where $M_U$ is the diagonal mass matrix. Especially for the top quark we
can rewrite~(2.23) as follows~[5]:
\beq
L_Y^t\ =\ -g_W\ \frac{m_t}{4M_W}\
H^0_i\ \bar{t} [ Y_i(1-\gamma_5) +Y_i^\ast (1+\gamma_5)] t.
\eeq 
The Yukawa couplings $Y_i$ in~(2.24) are given by
\beq
Y_i\ \ =\ \ \frac{d_{2i}}{s_\beta}\ +\ id_{3i}\cot \beta .
\eeq 
It is instructive to express~(2.24) in an equivalent form~[5]
\bea
L_Y^t \ &=& \ -g_W\ \frac{m_t}{2M_W}\
H^0_i\ [\alpha_{i2}\bar{t}t\ -\ i\alpha_{i3}\cot \beta
\bar{t} \gamma_5 t], \\
&&\alpha_{i2}\ =\ \frac{d_{2i}}{s_\beta}\ ,\quad \alpha_{i3}\ =\ d_{3i} .
\eea 
{}From~(2.26) it is evident that $L_Y$ leads to $CP$ violation, since
$H^0_i$ couples simultaneously to $CP$-even ($\bar{t}t$) and $CP$-odd
($i\bar{t}\gamma_5t$) currents.\\

\section*{3.~{\em CP} violation in the process
{\boldmath $e^+e^- \to t\bar{t} \nu_e\bar{\nu}_e$ \unboldmath}}
\stepcounter{chapter}
\indent

First we will discuss the manifestation of $CP$ violation in the process
$W^+W^- \to t\bar{t}$ which must be understood as a subprocess of the
reaction $e^+e^- \to t\bar{t} \nu_e \bar{\nu}_e$. Numerical results for such
processes has been obtained by using the Effective Vector Boson
Approximation~($EVBA$)~[16] in ref.~[17].\\

Regarding the signal of $CP$ violation in this process we consider the
production of left- and right-handed $t\bar{t}$ pairs. Under the
assumption of $CP$ conservation we have
\beq
<t_L( \vec{k}_t), \bar{t}_L(-\vec{k}_t)|\, \hat{\rho}\, |t_L( \vec{k}_t),
\bar{t}_L(-\vec{k}_t)>\ =\ <t_R( \vec{k}_t),
\bar{t}_R(-\vec{k}_t)|\, \hat{\rho}\, |t_R( \vec{k}_t), \bar{t}_R(-\vec{k}_t)>
\eeq 
in the centre of mass of the $t\bar{t}$ system,
where $\hat{\rho}$ is the density-state operator defined as
\beq
\hat{\rho} \ =\ \sum\limits_{\lambda_+,\lambda_-}
{\cal T}|W^+_{\lambda_+}(\vec{k}_+), W^-_{\lambda_-}(-\vec{k}_+)>
<W^+_{\lambda_+}(\vec{k}_+), W^-_{\lambda_-}(-\vec{k}_+)|{\cal T}^\dagger .
\eeq 
In eq.~(3.2) $\lambda_\pm$ denote the polarizations of the $W^\pm$ bosons and
${\cal T}$ is the transition operator.
In order to be condition~(3.1) valid it is sufficient to have ${\cal T}=
{\cal T^{CP}}$, then $\hat{\rho}=\hat{\rho}^{CP}$.
As a consequence, a non-zero value of the asymmetry
\beq
\hat{a}_{CP}\ \ =\ \ \frac{\hat{\sigma}(W^+W^- \to t_L\bar{t}_L)\ -\
\hat{\sigma}(W^+W^- \to t_R\bar{t}_R)}{\hat{\sigma}_{tot}}
\eeq 
would signal $CP$ violation. We have in mind the production of
left- and right-handed fermions in a high-energy linear $e^+e^-$ collider
of 1--2~TeV centre of mass energy~[18], where the polarization of the
top quarks produced lies mainly along or opposite to their momenta. The above
proposal of searching for $CP$ signals has been extensively discussed in~[5,6].
In fact, it has been argued that the final state $t_L\bar{t}_L$ can be
distinguished experimentally from $t_R\bar{t}_R$ by looking at the energy
distribution of the charged leptons (i.e. $e,\mu$) that result from the
subsequent decays of the top quarks.\\

It is worth observing that only scalars and pseudoscalar currents lead to
the production of left- and right-handed $t\bar{t}$ pairs. Indeed the
currents $J_{H_i}^{\lambda \lambda'}$ ($\lambda \lambda' = LL,\ RR$) due
to (2.24) read
\bea
J_{H_i}^{LL}\ & = &\ Y_i^\ast\ \bar{\mbox{u}}_L (1+\gamma_5) \mbox{v}_L ,\\
J_{H_i}^{RR}\ & = &\ Y_i \ \bar{\mbox{u}}_R (1-\gamma_5) \mbox{v}_R .
\eea 
The above two currents (i.e.~eqs (3.3) and (3.4)) imply a difference
in the amplitudes ${\cal A}(W^+W^- \to t_{\lambda} \bar{t}_{\lambda'})$
for $\lambda \lambda' = LL, RR$. This is, however, not enough to survive
this difference in the squared matrix element. As is well known, an
additional dynamical phase (beside the complex couplings $Y_i$) is required.
Usually, such phases arise from the calculation of absorptive parts
of higher order diagrams~[11]. Here, we do that in an effective way by
introducing complex $BW$ propagators for the Higgs bosons similar to~[12].
We, of course, assume that at least one of the Higgs masses is bigger than
$2m_t$. In general, it is known~[13,14] that one must consider
corrections due to final (initial) state
interactions  when discussing genuine $CP$-odd effects.
In other words, scatterings going to intermediate states
that are equal to final (initial) states will not contribute to the
$CP$-violating parameter $\hat{a}_{CP}$.
These additional constraints, the so-called
$CPT$ constraints, will be taken into account in the next section.
Here we simply quote the result for the squared matrix element using the
effective approach of $BW$~propagators. Defining
\beq
\tilde{Y}_i\ =\ Y_i\ g_{H_iWW}\ , \qquad \alpha_W\ =\ \frac{g_W^2}{4\pi} \, ,
\eeq 
we obtain, for example, for the dominant subprocess
$W^+_LW^-_L \to t_L\bar{t}_L,\ t_R\bar{t}_R$
via Higgs exchanges
\bea
|{\cal A}_{LL,RR}|^2 &=& 3\alpha^2_W \pi^2 \frac{m^2_t}{M^2_W}\
\frac{s^2_h(s_h-2m^2_t)}{M^2_W}\ \Bigg[ \sum\limits_i^{n_H}\
\frac{|\tilde{Y}_i|^2}{(s_h-M_i^2)^2+M_i^2\Gamma_i^2} \nonumber\\
&+&\ 2\ \sum\limits_{i<j}^{n_H}\ \mbox{Re}(\tilde{Y}_i^\ast \tilde{Y}_j)\
\frac{(s_h-M_i^2)(s_h-M_j^2)\ +\ M_iM_j\Gamma_i\Gamma_j}
{[(s_h-M_i^2)^2+M_i^2\Gamma_i^2]\ [(s_h-M_j^2)^2+M_j^2\Gamma_j^2]}
\nonumber\\
&\mp& \ 2\ \sum\limits_{i<j}^{n_H} \mbox{Im}(\tilde{Y}_i^\ast \tilde{Y}_j)\
\frac{(s_h-M_i^2)M_j\Gamma_j\ -\ (s_h-M_j^2)M_i\Gamma_i}
{[(s_h-M_i^2)^2+M_i^2\Gamma_i^2]\ [(s_h-M_j^2)^2+M_j^2\Gamma_j^2]}
\ \Bigg] .
\eea 
In eq.~(3.7) $n_H$ is the number of neutral Higgses (the result is then
valid for multi-Higgs models) and $s_h = k^2_h = (k_t + k_{\bar{t}} )^2$.
We are now in the position to define the physical $CP$ asymmetry
parameter $A_{CP}$ as follows:
\bea
A_{CP}\ &=& \ \frac{\Delta\sigma_{CP}(\ e^+e^- \ \to\  \nu_e\bar{\nu}_e
t\bar{t}\ )}{\sigma_{tot}(\ e^+e^- \ \to \ \nu_e\bar{\nu}_e t\bar{t}\ )}\, ,\\
\Delta\sigma_{CP}\ & = &\ \sigma(e^+e^- \to \nu_e\bar{\nu}_e t_L\bar{t}_L)
\ -\ \sigma(e^+e^- \to \nu_e\bar{\nu}_e t_R\bar{t}_R) .
\eea 
The result~(3.7) merely shows that $CP$ violation originates from the
subprocess $W^+W^- \to H^{0\ast}_i \to t\bar{t}$.
After considering the $CPT$ constraints mentioned
above, we  can exactly calculate $\Delta \sigma_{CP}$ with the help of
the Feynman graphs shown in fig.~1.
This can be done quite easily if the full matrix element squared has the
factorization property
\beq
\overline{|<t_\lambda\bar{t}_\lambda\nu_e\bar{\nu}_e|{\cal T}|
e^+e^->|^2}\ =\ F(s_h, M_{H_i})\
\overline{<\nu_e\bar{\nu}_e H_i^\ast|{\cal T}|e^+e^->|^2} ,
\eeq 
where $H_i^\ast$ denotes the $i$th virtual Higgs field with mass $\sqrt{s_h}$.
Then, the phase-space integral $R_4$ for this particular $2\to 4$ process can
be expressed as follows~[19]:
\beq
R_4(s_{tot})\ =\ \frac{\pi^2}{16s_{tot}} \ \int\limits_{4m_t^2}^{s_{tot}}
ds_h \int\limits_{\Gamma_3(s_h)} \frac{ds_2dt_1dt_2ds_1}{\Delta_4^{1/2}}
\ \frac{\lambda^{1/2}(s_h,m^2_t,m^2_t)}{2s_h}\ ,
\eeq 
where $s_i$ and $t_i$ are Mandelstam variables defined by
\bea
s_1\ &=&\ (k_\nu \ +\ k_h)^2, \nonumber\\
s_2\ &=&\ (k_{\bar{\nu}} \ +\ k_h )^2, \nonumber\\
t_1\ &=&\ (k_\nu \ -\ k_{e^-})^2, \nonumber\\
t_2\ &=&\ (k_{\bar{\nu}} \ -\ k_{e^+})^2 .
\eea 
The phase-space boundaries $s_i^\pm$ and $t_i^\pm$ are calculated
to be
\bea
s_2^\pm &=&\ \frac{1}{2}\ [(s_{tot} +s_h)\ \pm\ (s_{tot}-s_h)], \nonumber\\
t_1^\pm &= &\ -\frac{1}{2}\ [s_{tot}-s_2)\ \mp\ (s_{tot}-s_2)], \nonumber\\
t_2^\pm &=&\ -\frac{1}{2s_2}[(s_2-t_1)(s_2-s_h)\ \mp\ (s_2-t_1)(s_2-s_h)] ,
\nonumber\\
s_1^\pm &=&\ s_{tot}\ -\ \frac{1}{(s_2-t_1)^2} \left( \ D\ \mp\ 2\sqrt{F}
\ \right)
\eea 
with
\bea
D\ &=&\ (s_2-t_1)[s_{tot}(s_2-s_h)\ -\ t_2(s_{tot}+s_2)]\ +\ 2s_{tot}s_2t_2\ ,
\nonumber\\
F\ &=&\ s_{tot}\ s_2 \prod\limits_{i=1,2}\ (t_i-t_i^+) (t_i-t_i^-)\ ,
\nonumber\\
\Delta_4 &=&\frac{1}{16} (s_2-t_1)^2(s_1^+-s_1)(s_1-s_1^-)\ , \nonumber\\
\lambda (x,y,z)&=& (x-y-z)^2-4yz .
\eea 

\newpage

\section*{4.~{\em CPT} constraints}
\stepcounter{chapter}
\indent

In the following we will assume for simplicity that there is only one
intermediate Higgs state which is heavy and
we choose to be $H_h \equiv H^0_3$.\\

Consider the amplitude ${\cal T}$ of the diagrams depicted in fig.~1.
Since we are interested in $CP$-odd observables that are sensitive to the
Higgs width $\Gamma_h$, we do not calculate the dispersive parts of the
one-loop corrections.
Thus, we can list the
different contributing amplitudes to $\Delta\sigma_{CP}$ as follows:
\bea
{\cal T}_A \ &=& \ V_W\ \frac{g_{H_iWW}\ Y_j^\ast}{s_h-M_i^2}\ V_{t_L}\
\left( \ \delta_{H_iH_j}\  -\  i\mbox{Im}(\Pi_{H_iH_j})\ \frac{1}{s_h-M_j^2}
\ \right) ,  \\[0.5cm]
{\cal T}_B\
&=&\ V_W\ \frac{g_{H_hWW}\ Y_h^\ast}{s_h-M_h^2+iM_h\Gamma_h}\ V_{t_L}\
\left( \ 1\ -\ \frac{i\mbox{Im}(\Pi_{H_hH_h})\ -\ i M_h\Gamma_h}
{s_h-M_h^2+iM_h\Gamma_h}\ \right) , \\[0.5cm]
{\cal T}_C\ &=&\ V_W\ \frac{g_{H_hWW}}{s_h-M_h^2+iM_h\Gamma_h}\
i\mbox{Im}(\Pi_{H_hH_i})\ \frac{Y_i^\ast}{s_h-M_i^2}\ V_{t_L} ,\\[0.5cm]
{\cal T}_D\ &=&\ V_W\ \frac{g_{H_iWW}}{s_h-M_i^2}\ i\mbox{Im}(\Pi_{H_iH_h})\
\frac{Y_h^\ast}{s_h-M_h^2+iM_h\Gamma_h} \ V_{t_L} ,
\eea 
In addition, non-resonant contributions from $Z^0$-, $\gamma$- and $b$-quark
exchange graphs should also be considered, as well as absorptive parts of
vertex corrections and box graphs.
In eqs~(4.1)-(4.4) we sum over those
repeated indices $i$,$j$, which run over the non-resonant Higgs
states; $i,j=1,2$. $V_W$ is the amplitude of  $e^+e^- \to \nu_e
\bar{\nu}_e H^{0\ast}$ and $V_{t_L}$ the vertex function for
$H^{0\ast} \to t_L \bar{t}_L$. Then, in the
Born approximation we easily find that
\bea
\sum V^{(0)}_WV^{(0)\dagger}_W\ &=&\ 16 \pi^3 \alpha^3_W M^2_W\
\frac{s_{tot}(s_{tot}-s_1-s_2+s_h)}{(t_1-M_W^2)^2(t_2-M_W^2)^2} ,\\
\sum V^{(0)}_{t_L}V^{(0)\dagger}_{t_L}\ =\
\sum V^{(0)}_{t_R}V^{(0)\dagger}_{t_R}\
&=&\
3\pi\alpha_W\ (s_h-2m_t^2)\ \frac{m_t^2}{M^2_W} .
\eea 
The reason why the $\mbox{Im\,}\Pi_{hh}(q^2)-M_h\Gamma_h$ appears in the
numerator of the second term of eq.~(4.2) is that we expand the full
propagator in the gauge-invariant complex pole $M^2_h-iM_h\Gamma_h$
as follows~[20]
\bea
\frac{1}{q^2-M^2_h+i\mbox{Im\,}\Pi_{hh}(q^2)}\ &=&\
\frac{1}{q^2-M^2_h+iM_h\Gamma_h}\left(\  1\ +\
\frac{i\mbox{Im\,}\Pi_{hh}(q^2) - iM_h\Gamma_h}{q^2-M^2_h+iM_h\Gamma_h}
\ \right)^{-1} \nonumber\\
&=&\ \frac{1}{q^2-M^2_h+iM_h\Gamma_h}\left(\  1\ -\
\frac{i\mbox{Im\,}\Pi_{hh}(q^2) - iM_h\Gamma_h}{q^2-M^2_h+iM_h\Gamma_h}
\ \right)\nonumber\\
&& +\ \ {\cal O}(g_W^4) \ .
\eea 
This also reflects the on-shell renormalization condition
\beq
i\mbox{Im\,}\Pi_{hh}^R(q^2)|_{q^2=M^2_h}\ =\ i\mbox{Im\,}\Pi_{hh}(q^2)
\ -\ iM_h\Gamma_h|_{q^2=M^2_h}\ =\ 0 ,
\eeq 
at one-loop level.
We have checked that the matrix element given by eqs~(4.1)--(4.4) respects
the $CPT$ symmetry  in the resonant as well as off-resonant region. Note
that the way in which $i\mbox{Im\,}\Pi_{hh}^R(q^2)$ enters eq.~(4.7) is
crucial to show $CPT$ invariance in the off-resonant region.
However, it has been argued by Sirlin and Stuart in~[20] that the naive
ansatz of $BW$ propagators can induce gauge depedence in the matrix element
${\cal T}$. They further claim that the residue of the transition element
${\cal T}$ with respect to the complex-pole position $M^2_h-iM_h\Gamma_h$
and the background terms are separately gauge invariant. In this respect,
keeping terms up to ${\cal O}(g^2_W)$ at the resonant point, we can rewrite
eqs~(4.1)-(4.4) in a more convenient form, i.e.
\bea
{\cal T}_A\ &=&\ V_W^{(0)}\ \frac{g_{{H_i}WW}\ Y_i^\ast}{M^2_h-M_i^2}
\ V_t^{(0)} , \\[0.5cm]
{\cal T}_B\ &=&\ \frac{g_{{H_h}WW}\ Y_h^\ast}{s_h-M^2_h+iM_h\Gamma_h}\
[\ V_W^{(0)}V_t^{(0)}\ +\ i\mbox{Im}(V^{(1)}_W(M^2_h))V_t^{(0)}\ +\
iV^{(0)}_W \mbox{Im}(V^{(1)}_t(M^2_h))\nonumber\\
&& -iV_W^{(0)}\mbox{Im}(\Pi_{hh}'(M^2_h))V^{(0)}_t\ ], \\[0.5cm]
{\cal T}_C\ &=&\ V^{(0)}_W\ \frac{g_{H_hWW}}{s_h-M_h^2+iM_h\Gamma_h}\
i\mbox{Im}(\Pi_{H_hH_i})\ \frac{Y_i^\ast}{M^2_h-M_i^2}\ V^{(0)}_t ,\\[0.5cm]
{\cal T}_D\ &=&\ V^{(0)}_W\ \frac{g_{H_iWW}}{M^2_h-M_i^2}\
i\mbox{Im}(\Pi_{H_iH_h})\ \frac{Y_h^\ast}{s_h-M_h^2+iM_h\Gamma_h} \ V^{(0)}_t ,
\eea 
and the remaining $Z$-, $\gamma$- and $b$-quark mediated tree graphs
can formally be written as
\beq
{\cal T}_{tree}\ =\ \frac{V^{Z(0)}_W V_t^{Z(0)}}{M^2_h-M^2_Z}\ +\
\frac{V^{\gamma(0)}_W V_t^{\gamma(0)}}{M^2_h}\ +\ {\cal T}_b(M^2_h),
\eeq 
where ${\cal T}_b$ indicates the b-quark exchange graph at tree level.
Box graphs and other non-resonant vertex corrections do not contribute
at ${\cal O}(g^2_W)$ accuracy. Thus, the sum of the amplitudes given by
eqs~(4.9)-(4.13)
\beq
{\cal T}\ = \ {\cal T}_A\ +\ {\cal T}_B\ +\ {\cal T}_C\ +\ {\cal T}_D\
+\ {\cal T}_{tree},
\eeq 
represents a gauge-invariant expression. However, some terms in
${\cal T}$ will not contribute to $\Delta\sigma_{CP}$. Since the initial
states $W^+W^-$ of the subprocess are unpolarized, it suffices to consider the
tree-level value for the vertex $H-W-W$, i.e.~$V_W^{(0)}$. We also omit
terms proportional to ${\cal O}((s_h-M_h^2)/s_h)$ in $|{\cal T}_{CP}|^2$,
i.e.~${\cal T}_D$ and Im$\Pi_{hh}'(M^2_h)$ drop out. A further simplification
occurs by assuming that $M_i={\cal O}(M)$ for $i=1,2$,
i.e.~we neglect the mass difference $M_{H_1}-M_{H_2}$ as compared to the centre
of mass energy $\sqrt{s_{tot}}$. Finally, the non-resonant
amplitude ${\cal T}_{tree}$ has been left out for simplicity. In general,
the $Z^0(\gamma)$ particle will effectively operate only as an extra off-shell
particle with mass $M_Z\simeq M$ (or zero) in the context of a multi-Higgs
doublet model.\\

We now proceed with the computation of the relevant absorptive parts of
self-energies  $\Pi_{H_iH_j} (M^2_h)$ and the vertex function
$V^{(1)}_t(M^2_h)$  by adopting the Feynman--'t Hooft gauge.
The advantage of this gauge is that one can have a
safe estimate of the vertex $V^{(1)}_t$ for quite heavy Higgs masses by
taking only into account the unphysical Goldstone bosons (see also fig.~2).
We find that such corrections are either
supressed by a factor $m^2_t/M^2_h$ (e.g.~fig.~2.a) or show a soft
logarithmic dependence, $\ln (M^2_h/M^2)$, when increasing
the heavy Higgs mass $M_h$ (e.g.~graphs shown in figs~2.b--2.g).
In a recent work~[21], similar $CP$-odd observables have been calculated
in the decay process $H_h \to t\bar{t}$. There,
the $CP$ parameter $A_{CP}$ shows a quadratic behavior, i.e.~$M^2_h/M^2$,
different from that which we obtain by computing the Feynman graphs in
fig.~2. The reason lies in the fact that crucial $H^\ast_i$-mediated
self-energy graphs like
$H_h\to (W^+W^-, Z^0Z^0)\to H^\ast_i \to t\bar{t}$ have been omitted in~[21],
with $(W^+W^-, Z^0Z^0)$ denoting on-shell intermediate states. These graphs,
however, will guarantee gauge invariance for the above process.
Since we want to study the role of enhanced Higgs interactions in $A_{CP}$,
i.e. keeping terms of ${\cal O}(g^2_W M^2_h/M^2_W)$, such
self-energies have been consistently
included in our theoretical considerations by going
to the Feynman-'t Hooft gauge.
We think that the resonant part of the Higgs-mediated amplitude
will give the dominant contribution to the $CP$ asymmetry $A_{CP}$ in our
fixed gauge (i.e.~$\xi_{W,Z}=1$),
in spite of the omission of the vertex corrections.
In fact, $A_{CP}$ will behave asymptotically according to the form
\beq
|A_{CP}|\ \sim\ \frac{{\bar{\Gamma}}_h}{M_h}\ , \qquad
\bar{\Gamma}_h\ =\ \Gamma_h\ -\ \Gamma (H_h \to WW)\ -\ \Gamma(H_h\to ZZ),
\eeq 
which increases quadratically with the heavy Higgs mass (i.e.~$M^2_h/M^2_W$).
Therefore, it is also important to notice that
in a multi-Higgs scenario the total decay width of the heavy Higgs
$\Gamma_h$ can generally approximate the reduced width
$\bar{\Gamma}_h$ entering eq.~(4.15). This observation supports the above
approach of considering an amplitude in the Feynman-'t Hooft gauge,
which is dominated by Higgs interactions.
We finally remark that such a specific choice of gauge
corresponds to the equivalence principle between longitudinal vector bosons
and Goldstone fields by taking the limit $g_W\to 0$ with $v$ fixed,
i.e.~$g_W/M_W=2/v$.\\

We can now make use of the optical theorem and write down the absorptive parts
of the self-energies in the following way:
\beq
\mbox{Im}\, \Pi_{i_1i_2}(i_1\to ab\to i_2)|_{q^2=s_h}\ =\ \frac{g_{i_1ab}}
{g_{H_hab}}\ \frac{g_{i_2ab}}{g_{H_hab}}\ M_h\Gamma_{h\to ab}(s_h) ,
\eeq 
where $g$'s are Higgs coupling constants.
Then, $\Delta\sigma_{CP}$ can be cast into the form:
\beq
\Delta\sigma_{CP}\ \simeq\ F_{CP}(M^2_h)\ \int
\mbox{dPS}\ \frac{(\sum V_WV_W^\dagger)\
(\sum V_tV_t^\dagger)}{(s_h-M_h^2)^2\ +\ M^2_h\Gamma_h^2}\ \
\frac{4M_h\Gamma_h}{M^2_h-M^2}
\eeq 
with
\beq
F_{CP}(M^2_h)\ \  =\ \ \sum\limits_{i\not= h}\ \mbox{Im}(Y_iY_h^\ast )\
\sum\limits_{ab}\ \ \frac{g_{H_hWW}}{g_{H_hab}}\
\mbox{Br}(H_h\to ab)\
 \  \mbox{det}\left(
\begin{array}{cc}
g_{H_iWW} & g_{H_iab} \\
g_{H_hWW} & g_{H_hab} \end{array} \right)\,  .
\eeq 
The above expression of $F_{CP}$ is convenient to check $CPT$ invariance.
Although the quantity $\mbox{Im}(Y_iY_h^\ast )$ is not alone a genuine
$CP$~indicator ($CP$ violation also manifests itself in the constants
$g_{H_iab}$) one can, however, check that $F_{CP}$ vanishes in the limit
of $CP$ invariance, i.e.~$\mbox{Im}\lambda_5 (v_1^\ast v_2)^2 \ \to\ 0$.
Using eq.~(3.11) and inserting (4.5) and (4.6) into (4.17) we arrive at
the final expression for $\Delta\sigma_{CP}$, i.e.
\bea
\Delta\sigma_{CP}\ &\simeq &\ F_{CP}\ \frac{3\alpha_W^4}{256\pi^2}\
\frac{m_t^2 M_h\Gamma_h}{s_{tot}}\ \int\limits_{4m^2_t}^{s_{tot}}ds_h
\ \frac{\lambda^{1/2}(s_h,m^2_t,m^2_t)}{s_h}\
\int\limits_{\Gamma_3(s_h)}\frac{ds_2dt_1dt_2ds_1}{\Delta_4^{1/2}}\
\nonumber\\
&& \cdot \frac{M^2_h-2m^2_t}{[(s_h-M_h^2)^2+M^2_h\Gamma^2_h]
(M^2_h-M^2)}\ \ \frac{s_{tot}-s_1-s_2+M^2_h}{(t_1-M_W^2)^2(t_2-M_W^2)^2}\, .
\eea 
Some comments concerning the function $F_{CP}$ are now in order. In general
one has
\beq
F_{CP}\ = \ \sum\limits_{i}\ F_{CP}^{i},\quad \mbox{with}\quad
i\in\ \{t \bar{t},\ WW,\ ZZ,\ H_iZ,\ H_iH_j, \dots \},
\eeq 
where $i$ stands for all intermediate states (including also Goldstone
bosons and ghost fields) that can come on shell.
{}From eq.~(4.9) one easily derives that
\beq
F_{CP}^{WW}\ \ =\ \ F_{CP}^{ZZ}\ \ =\ \ 0 ,
\eeq 
which is a consequence of $CPT$ invariance.
In eq.~(4.18) it is also important to notice that
intermediate states with weak couplings proportional to the initial
couplings will not give any contribution to $F_{CP}$,
i.e.~$F_{CP}^{G^+G^-}=0$, $F_{CP}^{ZZ}=0$ etc.
As far as the final state interactions is concerned, we observe
that in the self-energy function $\mbox{Im}\, \Pi_{i_1i_2}(i_1\to
t\bar{t}\to i_2)$ there is no interference between scalar $(t\bar{t})_S$
and pseudoscalar $(t\bar{t})_P$ parts of the Yukawa Lagrangian~(2.26).
Recall that exactly this interference led to $CP$ violation in our process.
Therefore we can write
\beq
F_{CP}^{t\bar{t}}\ =\ F_{CP}^{(t\bar{t} )_S}\ +\ F_{CP}^{(t\bar{t} )_P}
\eeq 
with
\bea
F_{CP}^{(t\bar{t} )_S}\ & = &\ -\ \frac{3\alpha_W}{8\Gamma_h}\
\frac{m^2_t}{M^2_W}\ M_h\ \left( \ 1\ -\ \frac{4m^2_t}{M^2_W}\ \right)^{3/2}\
\frac{\cot^2 \beta}{\sin^2 \beta} \nonumber\\
& & \cdot d_{23}d_{13}d_{33}(\cos\beta\ d_{13}\ +\ \sin\beta\ d_{23})
,\\[0.5cm]
F_{CP}^{(t\bar{t} )_P}\ &=&\ \frac{3\alpha_W}{8\Gamma_h}\ \frac{m^2_t}
{M^2_W}\ M_h\ \left( \ 1\ -\ \frac{4m^2_t}{M^2_W}\ \right)^{1/2}\
d_{33}\ \cot^3 \beta  \nonumber\\
& & \cdot ( \cos\beta\ d_{13}\ +\ \sin\beta\ d_{23})\ (1\ -\ d^2_{13}\ +\
\cot\beta\ d_{13}d_{23}) .
\eea 
For completeness we also give the $F_{CP}$ term for the decay channel
$H_h\to H_iZ^0$ (including $G^0$)
\bea
F_{CP}^{H_jZ^0}\ &=&\ \frac{\alpha_W}{16\Gamma_h}
\ \frac{M^2_h}{M^2_W}\ M_h\ \lambda^{1/2}\left( 1,\frac{M^2_Z}{M^2_h} ,
\frac{M^2}{M^2_h} \right) \ \Bigg[ \left(\ 1\ -\ \frac{M^2}{M^2_h}\
\right)^2
+\ \frac{M^2_Z}{M^2_h}\ \Bigg( \ \frac{M_Z^2}{M^2_h}\ \nonumber\\
&& -\ 2\ \frac{M^2}{M^2_h}\ -\ 2\ \Bigg) \ \Bigg]\
d_{33}\cot\beta\ (-s_\beta\ d_{1j}\ +\ c_\beta\
d_{2j})^2\ (c_\beta\ d_{13}\ +\ s_\beta\ d_{23})\ \nonumber\\
&& \cdot (1\ -\ d_{13}^2\ +\ \cot\beta\ d_{13}d_{23}).
\eea 
Note that other decay channels, like $H_h\to H_iH_j$, $H_h\to H^+H^-$,
are, in principle, also possible. To avoid excessive complication,
decay modes involving trilinear Higgs couplings have been not considered in our
analysis. These decay modes, which depend strongly on the Higgs potential of
the model under discussion, are also gauge invariant by themselves.
In any case, due to the large number of free parameters existing in such
multi-Higgs scenarios it is unlikely to expect that accidental
cancellations or some kind of fine-tuning effect will occur between these
additional decay channels and those which we have already computed
in $F_{CP}$ such that $A_{CP}$ vanishes.\\

We close this section by making some important remarks. As we have seen,
initial state interactions play a crucial role here.
In general, it is important to know the production mechanism of the
unstable particles when studying $CP$ violation in their decays.
In our case we have to calculate the whole scattering process mediated by
the $H_h$ particle and the other light Higgses, {\em and not only the decay of
the $H_h$ particle}, since $WW$ and $ZZ$
intermediate states will not contribute to $A_{CP}$ when Higgses are produced
by $WW$ or $ZZ$ fusion in $e^+e^-$ collisions. Finally, the $CP$-violating
scattering  processes discussed above are not $s$-channel suppressed
at very high energies and are therefore worth investigation.\\

\section*{5.~Numerical estimates and discussion}
\stepcounter{chapter}
\indent

It is not easy to give an accurate numerical value for the $CP$ asymmetry
defining in eqs~(3.8) and (3.9),
since too many parameters entering $F_{CP}$ in~(4.18)
are unknown. However, in table 1 we have presented some typical
numerical results. For definiteness, we have varied $M_h$ between
$400-1000$~GeV for different top masses, i.e.~$ m_t=$ 100, 120, 140~GeV.
In addition, we fix the light Higgs masses to be $M=100$~GeV.
Thus, for example, for $\sqrt{s_{tot}}=2$~TeV
we find numerically that
\beq
\frac{\Delta\sigma_{CP}}{F_{CP}} \ \sim \ (0.1\ -\ 0.8)\ 10^{-2}\
\ \mbox{pb}
\eeq 
whithin the range of parameters given above.
To estimate $A_{CP}$ it seems reasonable to compute
$\sigma_{tot}$, using Standard-Model values for the Higgs couplings.
Doing so, we can have a first
estimate for $\sigma_{tot}$~[19] and the magnitude of $A_{CP}$
\beq
\sigma_{tot}\ \sim \ {\cal O} (10^{-3}-10^{-2})\ \mbox{pb}, \qquad
A_{CP}\ \sim\ {\cal O}(0.1-0.8)\ F_{CP} .
\eeq 
{}From eq.~(5.2) we also see that the value of $A_{CP}$ depends crucially
on $F_{CP}$. Nevertheless, from eq.~(4.20), the natural range of values
for $F_{CP}$ can be constrained by the inequality
\beq
F_{CP} \ \ \leq \ \ \frac{\bar{\Gamma}_h }{\Gamma_h}
\eeq 
In our $CP$-violating two-Higgs doublet scenario, this would correspond
to $F_{CP} \leq 1/2$.
We must also note that the $CP$-odd observable constructed
in eqs~(3.8) and (3.9) is only sensitive to absorptive parts at the one-loop
level of the reaction. In other words, typical $CP$- and $CPT$-odd observables
like $<\vec{s}_t \vec{k}_t >$, $< \vec{s}_{\bar{t}} \vec{k}_{\bar{t}}>$
have to combine with the $CPT$-odd absorptive graphs in order to give
a $CPT$-invariant contribution to the squared matrix element.
$CP$-odd observables that appear at tree level (e.g.~triple products
of the form $<\vec{k}_{e^-}\cdot\ \vec{k}_t \times \vec{k}_{\bar{t}}>$)
and are even under $CPT$, can be discussed in a similar way as given in~[22].\\

Of course, analogous calculations can be performed for the process
$e^+e^- \to Z^\ast Z^\ast \to e^+e^- t\bar{t} $ (see also table 2).
This together with the process discussed in this
paper could increase the statistics by considering the inclusive reaction
$e^+e^-\to t\bar{t}X$, with $X= \nu \bar{\nu} , e^+e^-$. We note that
at the high energy $e^+e^-$ linear collider one has to worry about
$t\bar{t}$ production through beamstrahlung photons~[17]. This would
increase the $CP$-even background and therefore a machine design with
low beamstrahlung luminosity will be preferable~[17]. Furthermore,
the most optimistic value for $A_{CP}$, estimated by the $s$-channel
suppressed process $e^+e^- \to Z^\ast \to t\bar{t}$, turns out to be
quite small, i.e.~$A_{CP} \simeq 10^{-3}$~[23], as compared to $A_{CP}
\simeq 40\%$ in our case. On the other hand, the top-pair production
cross section through $e^+e^-$ annihilation is of the order of 40~fb at
$\sqrt{s_{tot}}=2$~TeV and is hence competitive with numerical estimates
of cross sections for the vector-boson
fusion processes as presented in table 1 and 2.\\

In principle, Higgs-mediated
$CP$-odd processes could also play a crucial role in $pp$ colliders. However,
the overhelming top-production rate via gluon fusion will probably make such
$CP$-odd tests more difficult.\\

In conclusion we have discussed in detail how the $CP$ violation of the
Higgs potential with two Higgs doublets manifests itself in the
scattering process $e^+e^-\to \nu_e \bar{\nu}_e t\bar{t}$.
Our analysis will be also valid for multi-Higgs doublet models. We have
calculated the $CP$-odd resonant amplitudes of this reaction, paying
special attention to constraints resulting from $CPT$ invariance.
Using an approach which is $CPT$ and gauge invariant, we have demonstrated
that $CP$ asymmetries induced by Higgs-width effects
in $t\bar{t}$ production can be quite sizeable,
$10\% - 40\%$, and could be probed at a high energy $e^+e^-$ collider.\\[1.5cm]
{\bf Acknowledgements.} We thank J.~Bernab\'eu, J.~G.~K\"orner
and G.~Cvetic for helpful discussions on the subject. M.~Lavelle is gratefully
acknowledged for a critical reading of the manuscript. The work of
A.P.~has been supported by a grant from the Postdoctoral Graduate College
of Mainz and the work of M.N.~by the BMFT under the grant no.~055DO9188.

\newpage

\centerline{\bf\Large Figure and Table Captions }
\vspace{1cm}
\newcounter{fig}
\begin{list}{\bf\rm Fig. \arabic{fig}:}{\usecounter{fig}
\labelwidth1.6cm \leftmargin2.5cm \labelsep0.4cm \itemsep0ex plus0.2ex}

\item Feynman graphs of the resonant amplitudes contributing to
$\Delta\sigma_{CP}(\ e^+e^-\  \to\  \nu_e\bar{\nu}_e t \bar{t}\ )$.\\

\item Typical diagrams,
which give the leading contribution in the Feynman--'t Hooft gauge
to the one-loop coupling  $H_h-t-\bar{t}$ in the limit of $M^2_h \gg 4m^2_t$,
according to the equivalence principle (see also text).

\end{list}

\newcounter{tab}
\begin{list}{\bf\rm Tab. \arabic{tab}:}{\usecounter{tab}
\labelwidth1.6cm \leftmargin2.5cm \labelsep0.4cm \itemsep0ex plus0.2ex}

\item  Numerical results of the cross section $\sigma_{tot}^H$ and the
$CP$-asymmetry parameter $A_{CP}$ for
the process $e^+e^-\to H^\ast_i \to \nu_e\bar{\nu}_e t \bar{t}\ $
at $\sqrt{s_{tot}}=2$~TeV.  We fix $M=100$~GeV.
For $\sigma_{tot}^H$, we have assumed the Standard-Model values of the
couplings, i.e.~$g_{HWW}=g_{Ht\bar{t}}=1$ in our notation.
The three columns indicated by $I$, $II$, $III$ stand for values of
$m_t =$ 100, 120, 140~GeV, respectively.\\

\item Numerical results of the cross section $\sigma_{tot}^H$ and
the $CP$-asymmetry parameter $A_{CP}$ for
the process $e^+e^-\to H^\ast_i \to e^+e^- t \bar{t}\ $
at $\sqrt{s_{tot}}=2$~TeV,
using the same values for the set of parameters given in table 1.

\end{list}

\newpage

\bigskip\bigskip\bigskip\bigskip\bigskip\bigskip
\centerline{\bf\Large Table 1}
\vspace{1.5cm}
\begin{tabular*}{14.85cm}{|r||ccc||ccc|}
\hline
 & & & & & & \\
 $M_h$ & & $\sigma^H_{tot}$
$[$pb$]$&  & & $A_{CP}/F_{CP}$ & \\
$[$GeV$]$ &\hspace{.9cm} $I$\hspace{.9cm} &\hspace{.8cm}
 $II$\hspace{.8cm} & \hspace{.7cm}$III$\hspace{.7cm}
&\hspace{.5cm} $I$\hspace{.5cm} &\hspace{.4cm}
 $II$\hspace{.4cm} & \hspace{.3cm}$III$\hspace{.3cm}$ $ \\
\hline\hline
&&&&&& \\
400  & 7.2 $10^{-2}$ & 8.9 $10^{-2}$ & 1.0 $10^{-2}$ & 0.09 & 0.09 & 0.08 \\
500  & 3.7 $10^{-2}$ & 4.9 $10^{-2}$ & 6.1 $10^{-2}$ & 0.15 & 0.14 & 0.13 \\
600  & 1.9 $10^{-2}$ & 2.7 $10^{-2}$ & 3.4 $10^{-2}$ & 0.22 & 0.21 & 0.20 \\
700  & 1.1 $10^{-2}$ & 1.5 $10^{-2}$ & 1.9 $10^{-2}$ & 0.29 & 0.28 & 0.26 \\
800  & 5.6 $10^{-3}$ & 7.8 $10^{-3}$ & 1.0 $10^{-2}$ & 0.43 & 0.41 & 0.39 \\
900  & 3.0 $10^{-3}$ & 4.2 $10^{-3}$ & 5.5 $10^{-3}$ & 0.59 & 0.57 & 0.55 \\
1000 & 1.6 $10^{-3}$ & 2.2 $10^{-3}$ & 2.9 $10^{-3}$ & 0.82 & 0.80 & 0.78 \\
&&&&&& \\
\hline
\end{tabular*}

\newpage

\bigskip\bigskip\bigskip\bigskip\bigskip\bigskip
\centerline{\bf\Large Table 2}
\vspace{1.5cm}
\begin{tabular*}{14.85cm}{|r||ccc||ccc|}
\hline
 & & & & & & \\
 $M_h$ & & $\sigma^H_{tot}$
$[$pb$]$&  & & $A_{CP}/F_{CP}$ & \\
$[$GeV$]$ &\hspace{.9cm} $I$\hspace{.9cm} &\hspace{.8cm}
 $II$\hspace{.8cm} & \hspace{.7cm}$III$\hspace{.7cm}
&\hspace{.5cm} $I$\hspace{.5cm} &\hspace{.4cm}
 $II$\hspace{.4cm} & \hspace{.3cm}$III$\hspace{.3cm}$ $ \\
\hline\hline
&&&&&& \\
400  & 4.5 $10^{-3}$ & 5.6 $10^{-3}$ & 6.3 $10^{-3}$ & 0.09 & 0.09 & 0.08 \\
500  & 2.4 $10^{-3}$ & 3.1 $10^{-3}$ & 3.8 $10^{-3}$ & 0.14 & 0.13 & 0.12 \\
600  & 1.2 $10^{-3}$ & 1.7 $10^{-3}$ & 2.1 $10^{-3}$ & 0.21 & 0.20 & 0.19 \\
700  & 6.6 $10^{-4}$ & 9.2 $10^{-4}$ & 1.2 $10^{-3}$ & 0.30 & 0.29 & 0.27 \\
800  & 3.6 $10^{-4}$ & 4.9 $10^{-4}$ & 6.4 $10^{-4}$ & 0.41 & 0.39 & 0.37 \\
900  & 1.9 $10^{-4}$ & 2.7 $10^{-4}$ & 3.5 $10^{-4}$ & 0.59 & 0.57 & 0.55 \\
1000 & 1.0 $10^{-4}$ & 1.4 $10^{-4}$ & 1.9 $10^{-4}$ & 0.80 & 0.78 & 0.75 \\
&&&&&& \\
\hline
\end{tabular*}

\end{document}